\documentclass{aa}  
\usepackage{natbib}
\bibpunct{(}{)}{;}{a}{}{,} 
\usepackage{graphicx}
\usepackage{adjustbox} 
\usepackage{subcaption}
\usepackage{verbatim}
\usepackage[varg]{txfonts}

\usepackage{setspace}

\usepackage{booktabs}
\usepackage{multirow}
\usepackage{mydblfloatfix} 
\usepackage{comment}
\usepackage{float}
\usepackage{tikz}
\usepackage{color}
\graphicspath{{./}{./figure/}}

\usepackage{xcolor}

\usepackage[colorlinks=true,     linkcolor=blue, citecolor=blue, filecolor=blue, urlcolor=blue]{hyperref}

\usetikzlibrary{arrows,positioning} 
\tikzset{
    >=stealth',
    punkt/.style={
           rectangle,
           rounded corners,
           draw=black, very thick,
           text width=6.5em,
           minimum height=2em,
           text centered},
    pil/.style={
           ->,
           thick,
           shorten <=2pt,
           shorten >=2pt,}
}

\newcommand\rectagular[1][red]{\begin{tikzpicture}
\draw [fill=red,red] (0.2,0.2) rectangle (0.3,0.3); 
\end{tikzpicture}
}

\begin{document}

     \title{Assessment of the near-Sun magnetic field of the 10 March 2022 coronal mass ejection observed by Solar Orbiter}
    \author{S.Koya\inst{1,3}
    \and S.Patsourakos\inst{1} 
    \and M.K Georgoulis\inst{2,4}
    \and A. Nindos\inst{1}
  }

\institute{Section of Astrogeophysics, Department of Physics, University of Ioannina, 45110, Greece
\email{s.koya@uoi.gr} 
\and Space Exploration Sector, Johns Hopkins Applied Physics Laboratory, Laurel, MD 20723, USA
\and Institute of Physics, University of M. Curie-Skłodowska, Pl. M. Curie-Skłodowskiej 1, 20-031 Lublin, Poland 
\and Research Center for Astronomy and Applied Mathematics, Academy of Athens, 11527 Athens, Greece
}

    
\date{Received date / Accepted date }

\abstract {}{We estimate the near-Sun axial magnetic field of a coronal mass ejection (CME) on 10 March 2022. Solar Orbiter's in situ measurements, 7.8 degrees east of the Sun-Earth line at 0.43 AU, provided a unique vantage point, along with the WIND measurements at 0.99 AU. We determine a single power-law index from near-Sun to L1, including in situ measurements from both vantage points.} {We tracked the temporal evolution of the instantaneous relative magnetic helicity of the source active region (AR), NOAA AR 12962. By estimating the helicity budget of the pre-and post-eruption AR, we estimated the helicity transported to the CME. Assuming a Lundquist flux-rope model and geometrical parameters obtained through the graduated cylindrical shell (GCS) CME forward modelling, we determined the CME axial magnetic field at a GCS-fitted height. Assuming a power-law variation of the axial magnetic field with heliocentric distance, we extrapolated the estimated near-Sun axial magnetic field to in situ measurements at 0.43 AU and 0.99 AU.}{The net helicity difference between the post-and pre-eruption AR is $(-7.1 \pm 1.2) \times  10^{41} \mathrm{Mx^{2}}$, which is assumed to be bodily transported to the CME. The estimated CME axial magnetic field at a near-Sun heliocentric distance of 0.03 AU is 2067 $\pm$ 405 nT. From 0.03 AU to L1, a single power-law falloff, including both vantage points at 0.43 AU and 0.99 AU, gives an index $-1.23 \pm 0.18$.} {We observed a significant decrease in the pre-eruptive AR helicity budget. Extending previous studies on inner-heliospheric intervals from 0.3 AU to $\sim$1 AU, referring to estimates from 0.03 AU to measurements at $\sim$1 AU.  Our findings indicate a less steep decline in the magnetic field strength with distance compared to previous studies, but they align with studies that include near-Sun in situ magnetic field measurements, such as from Parker Solar Probe.}

\titlerunning{Near-Sun axial magnetic field of 10 March 2022 CME}
\authorrunning {S.Koya, S.Patsourakos, M.K Georgoulis, A. Nindos}

\maketitle
\section{Introduction} \label{introduction}
Coronal mass ejections (CMEs) are energetic eruptions of plasma and magnetic fields from the solar atmosphere \citep{forbes2000, Chen2011, WebbHoward2012}. 
When detected in situ in interplanetary space, they are called interplanetary CMEs (ICMEs). Despite substantial aerodynamic drag and subsequent magnetic erosion during interplanetary (IP) propagation \citep{Gopal2001Dragmodel, Tappin2006Drag, Sachdeva2015Drag, Stamkos2023}, understanding the near-Sun magnetic field strength of CMEs is crucial as it plays a key role in shaping the early evolution of CME dynamics and the potential Earth impact at L1. This knowledge could be vital in space weather research, especially for Earth-directed CMEs. The intensity of the southward IP magnetic field associated with Earth-directed CMEs is crucial in determining the magnitude of the ensuing geomagnetic storms \citep{WuLepping2005}. In order to evaluate the geoeffectiveness of CMEs, only a few estimates of the near-Sun magnetic field are available, such as a few indirect methods \citep{Kunkel_2010, Savani2015} and some based on rare radio-emission configurations such as gyrosynchrotron emission from CME cores and Faraday rotation \citep{Bastin2001, JensenRusell2008, TunSamuelVaourlidas2013}. These diagnostic tools are unsuitable for routine use due to the lack of solar radio arrays with the necessary sensitivity and capabilities for continuous observations.

The significance of free (above the lower-level vacuum [potential] solution) and total magnetic energy in initiating CMEs is widely recognised \citep{priest_2014}. However, the precise role of magnetic helicity in the initiation of solar eruptions is still being investigated.
In active regions (ARs), the total magnetic energy encompasses both the stable potential energy and the dynamic non-potential (free) terms. The free magnetic energy in ARs results from the combination of magnetic flux emergence and the intricate movements of plasma, corroborated by emerging, electric current-carrying magnetic flux tubes \citep[e.g.][]{Leka1996ApJ}.
The release of free magnetic energy, believed to occur through intermittent episodes of magnetic reconnection, is the key reason for solar magnetic activity.  Meanwhile, magnetic helicity, which represents a measure of the twist, writhe and linkage of magnetic field lines, is also found essential in understanding solar eruptions \citep{berger1984rigorous}. Observational support for the importance of magnetic helicity in the initiation of solar eruptions includes works by \citet{Rust_Kumar96}, \citet{ nindos_andrews2004}, \citet{park2008, Park2010}, \citet{Nindos2012}, \citet{Liu2023}, \citet{Liokati_2022}, and \citet{Liokati2023}. Earlier analysis and modelling works have presented solar eruptions as a necessary outcome of the presence of magnetic helicity \citep{Low94,KumarandRrust1996}. There are multiple methods for computing the relative magnetic helicity (that is, the helicity due to non-potential magnetic fields) using observational data - a comparison of different methods can be found in \citet{Thalmann2021MagneticHE}. The combined knowledge of free magnetic energy and relative magnetic helicity may present a path for a more substantive understanding of magnetic configurations in the solar atmosphere. 

The magnetic helicity of an AR source 
can be indirectly used to understand the near-Sun magnetic field of the associated CME by linking it to an idealised magnetic flux rope structure. \citet{Patsourakos_2016} introduced a method based on the conservation of magnetic helicity during CMEs. The approach involves calculating eruption-related magnetic helicity from photospheric magnetic fields and employing the graduated cylindrical shell (GCS) model of \citet{Thernisien_2006} and \citet{Thernisien2009} on multi-viewpoint coronagraph observations such as from Solar Terrestrial Relationship Observatory \citep[STEREO;][]{STEREO2008} and Large Angle and Spectrometric Coronagraph \citep[LASCO;]{LASCO1992} on board the Solar and Heliospheric Observatory \citep[SOHO;][]{SOHOMISSION1995}.  The method utilises analytical relationships developed for various models, such as those by \citet{Dasso2006}, to determine the ICME's magnetic field magnitude. This magnitude is then extrapolated to various heliocentric distances up to 0.99 AU by assuming a self-similar, power-law decrease with radial distance. 

\citet{Patsourakos_2016} applied this methodology to the major geoeffective CME on 7 March 2012. The study utilised three relative helicity estimation methods: the helicity injection method from \citet{paritat2006}, the nonlinear force-free connectivity-based (CB) method by \citet{Georgoulis_2012}, and a nonlinear force-free volume calculation of helicity as described in \citet{Moraitis2014}. Using analytical expressions relating magnetic helicity to cylindrical flux rope configurations, the study estimated the near-Sun magnetic field at 13 $R_{\odot}$ in the range [0.01, 0.16] G. This magnetic field exhibited a steep radial falloff with a power-law index of -2.0 when compared to the magnetic fields of corresponding ICMEs at a distance of 0.99 AU from the Sun. 
In a subsequent study by \citet{PatsourakosPrametric}, a parametric analysis of the CB method was conducted to evaluate its robustness. The study used the input parameter distributions derived from observations to determine near-Sun and L1 magnetic fields for synthetic CMEs. It was found that the near-Sun CME magnetic fields at 10 $R_{\odot}$ ranged between 0.004 G and 0.02 G, which is comparable to, or higher than, the nominal magnetic field in the quiescent corona at the same distance. A power-law exponent $\alpha _B = -1.6 \pm 0.2$ was found to be most consistent with magnetic cloud (MC) field magnitudes at L1. In \citet{Patsourakos2017stellar}, this parametric study was extended to encompass a variety of proposed theoretical CME models, thereby further refining the understanding of radial field evolution in these events. The study also expanded to non-solar CME cases, particularly stars that host superflares. It aimed to determine magnetic fields of extreme stellar CMEs, knowledge that was later used even to define preliminary conditions for the habitability of exoplanets subjected to such CMEs  \citep{Samara_2021}. Nevertheless, it is crucial to validate this approach more extensively by including a broader range of events and conducting comprehensive comparisons with other methods. 
The current landscape, enriched by data acquired from missions such as Solar Orbiter(SolO) and Parker Solar Probe (PSP), provides more observational possibilities that could further facilitate this validation process.

\begin{figure}[h!t]
\resizebox{\hsize}{!}{\includegraphics{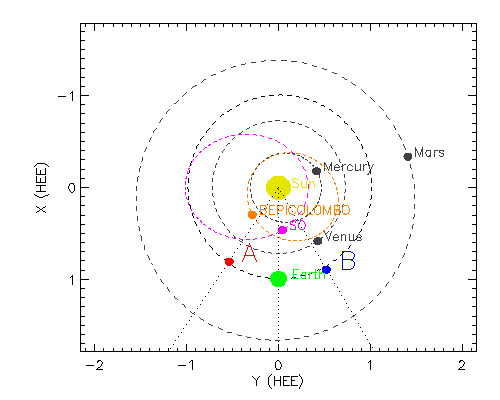}}
\caption{Spacecraft configuration of STEREO B (blue circle), STEREO-A (red circle), SolO ( magenta circle), Bepi Colombo (orange circle), Earth (green circle) and Sun (yellow) on 10 March 2022 at 00:00 UT in heliocentric Earth ecliptic (HEE) system. The dotted lines show the angular difference of STEREO A and B from the Sun-Earth line. Units are in AU.}
\label{spacecraftposiion}
\end{figure}

We present the estimation of the near-Sun axial magnetic field, coupled with a Sun-to-Earth analysis of the CME observed at L1 and by the SolO mission \citep{SolOMissionPaper} on 10 March 2022, employing the methodology proposed in \citet{Patsourakos_2016}. The novelty of this study is the advantageous SolO position, just 7.8 degrees east of the Sun-Earth line (see Fig.\ref{spacecraftposiion}), at a heliocentric distance of 0.43 AU along with the  L1 measurements by the WIND spacecraft \citep{WINDMAGNETICFIELD}. The two spacecraft measurements provided a rare opportunity for robust cross-validation of the ICME magnetic field at different heliocentric distances. The specialised spacecraft configuration draws particular attention to this event, such as in 
\citet{Jackson2023}, who utilised 3D reconstruction modelling from interplanetary scintillation forecast techniques to analyse this CME and its potential impact. \citet{Laker2023} used real-time measurements from SolO to predict the CME's arrival time and magnetic structure more than a day before it arrived at Earth. Furthermore, \citet{Zhuang2024} investigated the correspondence between remote observations and in situ measurements of this CME using data from SolO and STEREO/heliospheric imagers. This was achieved by comparing CME properties derived from both techniques, such as size and radial expansion, within 0.5 AU. 
Our work contributes to a better understanding of the event by focusing on the magnetic field evolution of this CME from the near-Sun to Earth, using the source active region's helicity budget and SolO measurements taken halfway between the Sun and Earth.
This work also paves the way for further works, particularly in cases in which SolO is similarly positioned or aligned with other spacecrafts, such as PSP.

The article is organised as follows: Section \ref{sec:Observations} summarises remote-sensing and in situ measurements of the 10 March 2022 CME. Section \ref{Methodology} describes the methodology used to estimate the near-Sun axial magnetic field of the CME. The extrapolation of the estimated magnetic field from near-Sun to 0.43 AU and 0.99 AU is then presented in Section \ref{Results}. Then, Section \ref{discussions& Conclusions} compares our results with previous studies and discusses advances in understanding and potential future research in this direction. The article includes two appendices: Appendix \ref{data gap MAG/Solo} explains the methodology used to address the data gap seen in SolO magnetic field measurements during the ICME, while Appendix \ref{monte-carlo procedure} describes the Monte Carlo approach used to estimate uncertainties in our near-Sun axial magnetic field estimation.

\section{Observations}\label{sec:Observations}
\subsection{Solar source active region}

\begin{figure*}

    \includegraphics[width=\textwidth]{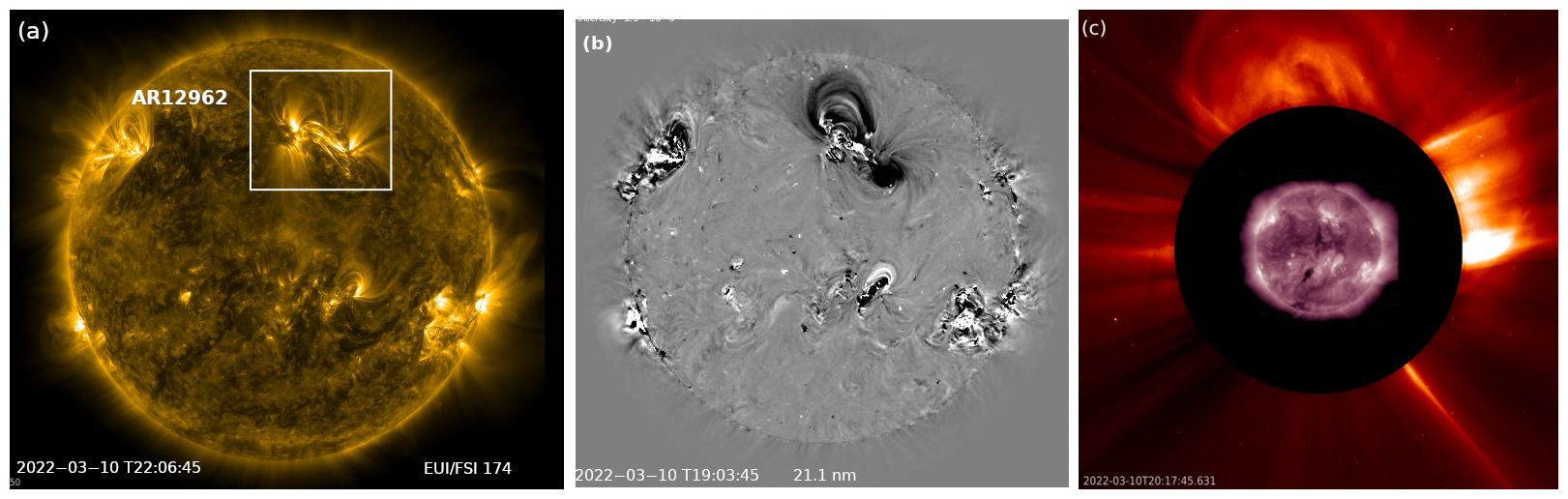}
    \caption{ (a) Full-disk image by the EUI/FSI 174 \AA~imager. The source AR 12962 is enclosed by the white box. The image shows a post-eruption arcade. (b) Full-disk base difference image at near dimming peak at 19:03 UT by SDO/AIA 211 \AA~channel with respect to the image at 17:00UT (c) Partial halo CME observed by SOHO/LASCO C2 on 10 March at 20:17 UT with the overlaid full-disk image by SDO/AIA 211 \AA~channel.}
  \label{CME remote sensing 3 images}
\end{figure*}

On 10 March 2022, the Atmospheric Imaging Assembly \citep[AIA;][]{AIA2012} telescope on board the Solar Dynamics Observatory(SDO) \citep{SDO2012}, SWAP/PROBA-II \citep{SWAPPROBA_Seaton2013, SWAPPROBHalain2013} and Extreme Ultraviolet Imager (EUI)/ Full Sun Imager (FSI) \citep{EUI_Solo_2020} 174 \AA$\;$ channel on board SolO observed signs of the eruption from NOAA AR 12962 using full-disk Extreme Ultraviolet (EUV) images (see Fig.\ref{CME remote sensing 3 images}a). Key observations included loop expansion starting at 17:10 UT on March 10 in the EUI/FSI 
174 \AA~channel, 
and coronal dimmings start time at 16:22 UT in AIA 211 \AA$\;$ channel, followed with a dimming peak time at 18:58 UT in the AIA 211 \AA~channel (see Fig.\ref{CME remote sensing 3 images}b). This is followed by the appearance of a post-eruption arcade, visible in the AIA 304 \AA~channel and, 30 minutes later, in the 174\AA~AIA channel. An accompanying Geostationary Operational Environmental Satellites (GOES) C2.8 class flare was observed with the start time around 19:00 UT and peak at 20:33 UT. The LASCO/SOHO later observed this event as a partial halo CME at around 19:00 UT (see Fig.\ref{CME remote sensing 3 images}c) along with the detection by the Sun–Earth Connection Coronal and Heliospheric Investigation (SECCHI) aboard the STEREO spacecraft \citep{SECHII2008, STEREO2008} as a limb CME.

\subsection{CME- ICME connection}
Here, we show evidence of the CME-ICME connection for this event.  Based on the positioning of the spacecraft (see Fig.\ref{spacecraftposiion}) and AR location along with the CME's tilt, orientation and angular width derived from GCS modelling ( see Sect.\ref{Results-forward modelling}),  we infer that the CME observed in full-disk and coronagraphic images on 10 March 2022 likely impacted both SolO and WIND. To further confirm this, we followed the procedure described in \citet{shooting_back_cme_Zhang2007}: by assuming a constant propagation speed, we find the difference between the estimated time of arrival and the actual arrival time both at SolO at 0.43 AU and at WIND at 0.99 AU. Table \ref{CME-ICME connection table and resulting time of arrival  differences} lists the results using various detection tools and databases. We used speed estimations from various CME catalogues such as CACTUS  \footnote{\url{https://www.sidc.be/cactus/catalog.php}}, Solar Eruptive Event Detecting System (SEEDS) / LASCO and SEEDS/STEREO \footnote{\url{http://spaceweather.gmu.edu/seeds/}}, LASCO CME catalogue \footnote{\url{https://cdaw.gsfc.nasa.gov/CME_list/}}, The Space Weather Database Of Notifications, Knowledge, Information (DONKI) catalogue \footnote{\url{https://kauai.ccmc.gsfc.nasa.gov/DONKI/}} and  ARRival CATalog (ARRCAT) \footnote{\url{https://helioforecast.space/arrcat}}. The ARRCAT catalogue estimates CME speed by modelling the HELiospheric Cataloguing, Analysis and Techniques Service (HELCATS)\footnote{\url{https://www.helcats-fp7.eu/catalogues/wp3_cat.html}} CMEs time-elongation tracks with the SSEF30 model \citep{Davisssf02012, Mostl2014CONNECTINGSD} single-spacecraft ﬁtting technique.  In Table \ref{CME-ICME connection table and resulting time of arrival  differences}, the time of arrival (ToA) estimate, which is later than the actual arrival time (this is discussed in detail in Sect.\ref{in situ observations}), is indicated as positive, and an earlier ToA than the actual is indicated as negative.  We also included the estimated ToA from the GCS  speed estimation from our GCS fitting using a linear fit of height-time data in consecutive GCS images ( see Fig.\ref{GCSFitted} ). Further evidence for our CME-ICME connection results from the GCS model application is also discussed in Sect.\ref{Results-forward modelling}. 

\begin{table*}[ht]
\caption{CME -ICME connection table;a comparison of the estimated and actual arrival time of ICME at SolO and WIND.
}
\label{CME-ICME connection table and resulting time of arrival  differences}
\centering

\begin{tabular}{|l|l|l|l|l|l|l|}
\hline
 Detection tool & Time stamp used  & Speed & Estimated time & Estimated & $T_{\text{est.SolO}}$ - & $T_{\text{est.WIND}}$ - \\
 & in the calculations of  & (km/s) & of arrival & time of arrival & $T_{\text{actual}}$ & $T_{\text{actual}}$ \\

 & the ToAs, 2022-03-10 & & at 0.43 AU& at 0.99 AU &  & \\
\hline

CACTUS& 18:48  &  385 & 2022-03-12 17:21  & 2022-03-15 04:51  &0D 21H 29M & 1D 18H 44M \\

SEEDS/LASCO& 19:00 & 335 & 2022-03-13 00:29 & 2022-03-15 20:53 &  1D 04H 37M & 1D 18H 44M
 \\

SEEDS/STEREO& 20:53 & 657 & 2022-03-12 00:09& 2022-03-13 11:02 & 0D 04H 16M& 0D 00H 55M \\

LASCO & 18:48 & 742 & 2022-03-11 18:57 & 2022-03-13 01:49 &0D 0H 56M & 0D -09H 58M
 \\

DONKI & 19:47 & 677 & 2022-03-11 22:15 & 2022-03-13 08:05 & 0D 02H 22M  & 0D -02H 42M 
\\

ARRCAT& 18:14 & 677 & 2022-03-11 20:56 & 2022-03-13 06:32 & 0D 00H 49M 
& 
0D -03H 01M 
 \\

\hline
\end{tabular}
\end{table*}

Table \ref{CME-ICME connection table and resulting time of arrival  differences} shows that the maximum obtained difference between the estimated and the actual ToA at SolO is 1 day 4 hours 37 minutes when using the SEEDS/LASCO speed of 335 km/s, and a minimum is 49 min when using the ARRCAT speed estimation. The corresponding difference in time estimations at 0.99 AU also shows a similar trend, with a maximum difference of 1 day 18 hours 44 minutes with the SEEDS/LASCO speed estimate and 03 hours 01 minutes early estimation from the ARRCAT catalogue. The large difference between actual and estimated time of arrival, calculated using LASCO speed estimation, is largely caused by the projection effects of the CME onto a two-dimensional plane. It is important to note that the two-dimensional projection of CMEs does not accurately represent their true three-dimensional propagation shape and structure. Studies by \citet{Temmer2009}, \citet{HOWARD2008} and \citet{Vrnak2007ProjectionEI} have shown that the velocity and angular width of CMEs are particularly affected by the projection effect and the angle of observation. However, these effects are minimised for limb CMEs \citep{Burkepile2004, CID2012}. This CME appears as a partial halo CME in the SOHO/LASCO field of view, while it is a limb CME in the STEREO field of view. As viewed from the Earth the source of the CME was not far from the central meridian. Hence, the speeds inferred by LASCO images may not be close to the actual speed. 

To summarise, the differences between the predicted and observed ToAs, as presented in Table \ref{CME-ICME connection table and resulting time of arrival  differences}, range from a maximum of 1 day, 18 hours, 44 minutes (according to the CACTUS catalogue) to a minimum of 49 minutes (according to the ARRCAT catalogue). Considering the aforementioned caveats, as well as the uncertainties introduced by assuming constant speed propagation, which could account for these differences, we conclude that the CME observed in the full disk EUV and coronagraphic images on 10 March 2022 has indeed reached both the SolO at a distance of 0.43 AU and then the WIND spacecraft at 0.99 AU. This CME-ICME identification is in agreement with recent studies of this event, such as \citet{Laker2023},\citet{SoLo_first},\citet{Jackson2023} and \citet{Zhuang2024}.

\subsection{In situ ICME observations at 0.43 AU and 0.99 AU}\label{in situ observations}

\begin{figure*}[ht]
\centerline{
\includegraphics[width=17cm]{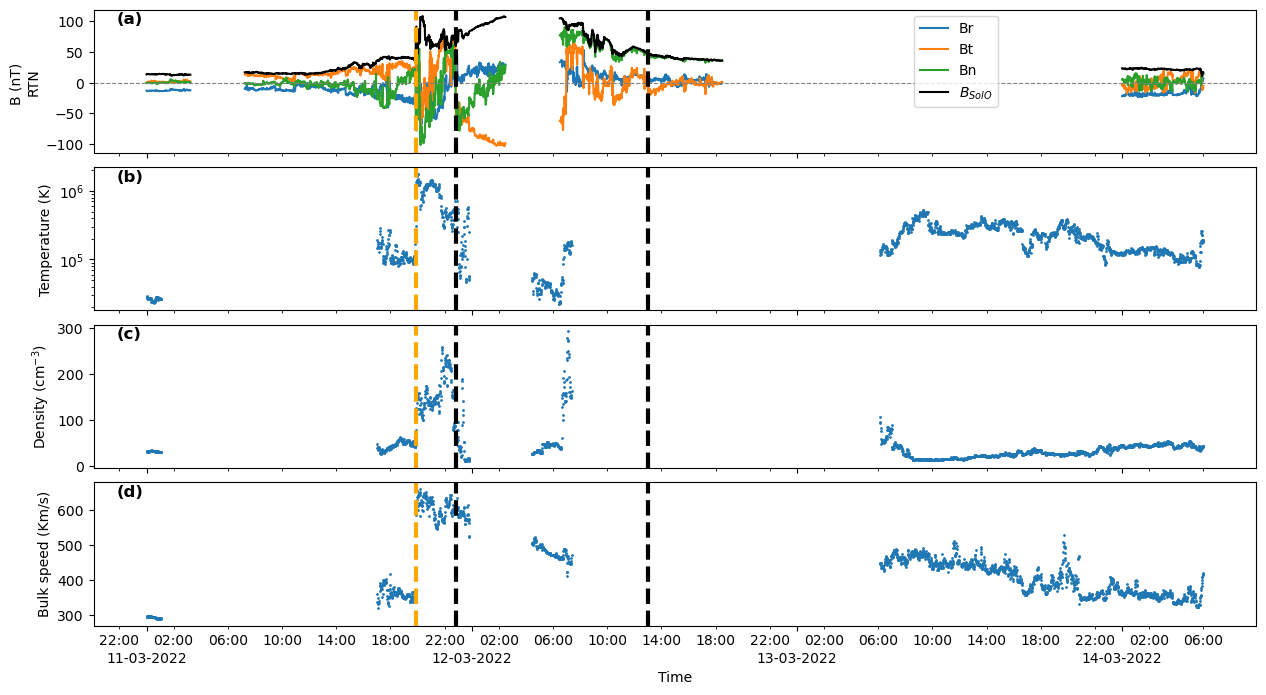}
}
\caption{Solar Orbiter in situ observations at 0.43 AU: a) magnetic ﬁeld components in the RTN system and the magnitude of the magnetic field, b) proton temperature, c) proton density, and d) solar wind bulk velocity. The vertical dashed line marks the shock's arrival (orange), and the two black vertical dashed lines represent ICME start and end times, respectively. 
}
\label{solo insitu}
\end{figure*}

We now discuss the in situ ICME measurements associated with the CME by SolO at 0.43 AU and WIND at 0.99 AU. Fig.\ref{solo insitu} presents solar wind plasma and magnetic field measuremnts obtained from the Proton and Alpha Sensor (PAS) \citep{SWASolo} and magnetometer (MAG) \citep{MAGSolo2020} instruments on board SolO. We analysed SolO Level 2 Magnetometer Data in Radial Tangential Normal (RTN) Coordinates in Normal Mode with a 1-minute cadence collected from 11 March 2022, 00:00 UT, to 14 March 2022, 06:00 UT. The top panel (Fig.\ref{solo insitu}a) illustrates the magnetic field vector components $B_r$, $B_t$, $B_n$ in the RTN system along with the magnitude of the magnetic field ($B_{SolO}$). The figure also includes plasma data on proton temperature (Fig.\ref{solo insitu}b), density (Fig.\ref{solo insitu}c), and solar wind bulk velocity (Fig.\ref{solo insitu}d) Level 2 data with a 1-minute cadence. These temporal variations of  ICME plasma parameters and magnetic field components are used for identifying the shock, sheath, and ICME boundaries. The dashed vertical lines in each panel mark the identified shock arrival (yellow), ICME start and end times (black).

On approximately 11 March 2022, around 19:53 UT, a marked jump in magnetic field magnitude, proton temperature, proton density, and solar wind velocity indicated the arrival of the shock at SolO. The shock is followed by a sheath region characterised by signiﬁcant ﬂuctuations in all related physical parameters. Around March 11, 22:47 UT marks the ICME start time, which is characterised by more gradual changes in these parameters. An intriguing observation occurred on 11 March around 22:48 UT, when magnetic field components started to exhibit significant rotations. A smooth rotation of both the $B_t$ and $B_n$ components was observed. The $B_t$ component transitioned from a positive to a negative value, while the $B_n$ component transitioned from a negative to a positive value. This rotation indicated a progression towards the maximum field of an ICME. However, there is a discontinuity due to a data gap. Subsequently, it was observed that the $B_t$ component reached a negative peak, and the $B_n$ component reached a positive peak. This rotation suggests the possible existence of an MC structure within the ICME \citep{Burlaga1988}. 

There is a significant data gap in both magnetic field and PAS measurements during the sheath region pass (from 23:49 UT to 04:36 UT) and the ICME region (from 07:21 UT to 06:10 UT) that obscured further changes, making it challenging to establish definitive boundaries for the ICME. Nonetheless, by analysing magnetic field components and solar wind parameters until March 14 around 06:00 UT (when the variation exhibits a more gradual trend in magnetic field components) and comparing them to the ambient solar wind, we determined ICME end time to be around March 12 12:00 UT. To determine the maximum magnetic field strength of the ICME ($B_{0_{SolO}}$) at SolO and its potential uncertainty, a technique for resolving data gaps (as described in Appendix \ref{data gap MAG/Solo}) was utilised. This led to an estimation of $B_{0_{SolO}}$=109.72 $\pm$ 20.27 nT. 

\begin{figure*}[ht]
\centerline{
\includegraphics[width=17cm]{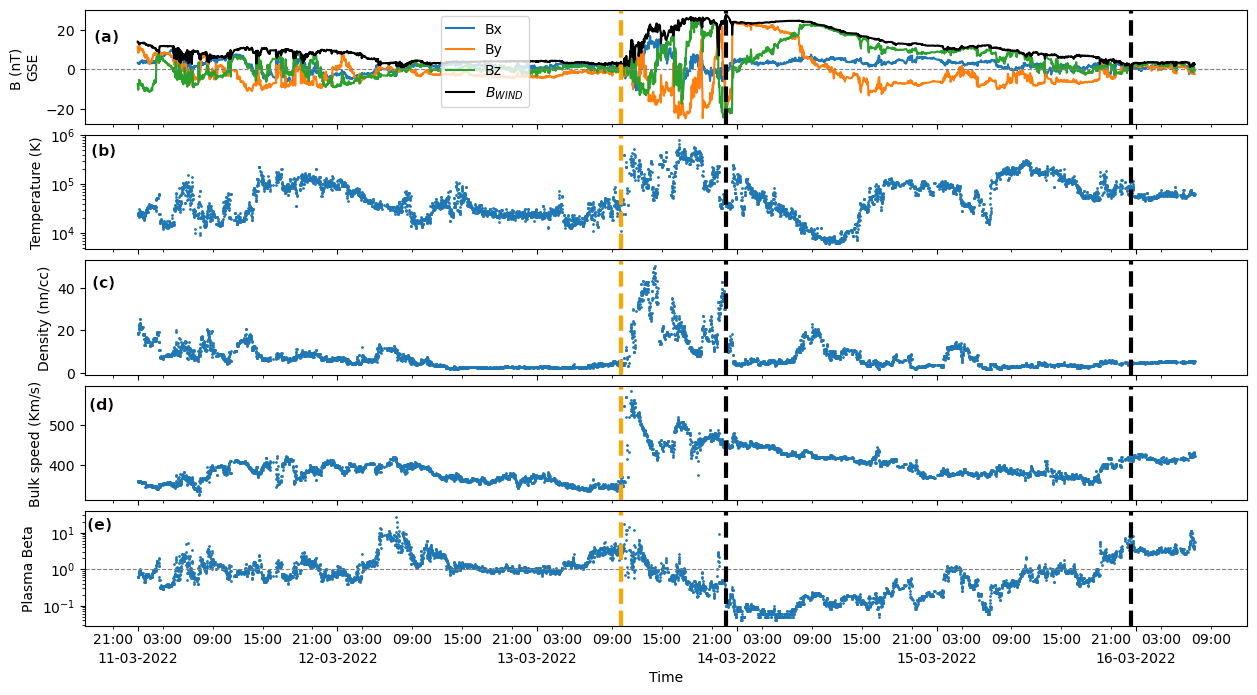}
}
\caption{Wind L1 in situ observations (from top to bottom): a) magnetic-ﬁeld components in the GSE system with the magnitude of the magnetic field, b) proton temperature, c) proton density, d) solar wind bulk velocity, and e) proton plasma $\beta$. The vertical orange dashed line marks the shock's arrival, and the two black vertical dashed lines represent ICME start and end times.}
\label{WINDinsitu}
\end{figure*} 
Similarly, Fig.\ref{WINDinsitu} shows solar wind plasma measurements by SWE \citep{WINDPLASMA} and magnetic field measurements by MFI \citep{WINDMAGNETICFIELD} from the WIND spacecraft at 0.99 AU. These are Level 2 data with a 1-minute cadence between 11 March 2022 at 02:00 UT and 16 March 2022 at 07:00 UT. 
Fig.\ref{WINDinsitu}a displays the magnetic field vector components $B_x$, $B_y$ and $B_z$ all in the GSE (Geocentric Solar Ecliptic)) coordinates along with the magnitude of the magnetic field ($B_{WIND}$). 
Subsequently, proton temperature (Fig.\ref{WINDinsitu}b), proton density (Fig.\ref{WINDinsitu}c), solar wind bulk speed (Fig.\ref{WINDinsitu}d), and plasma beta (Fig.\ref{WINDinsitu}e), all with a 1-minute cadence, are shown. Similar to SolO in situ data, these plasma parameters and the magnetic field variation are used to define the shock, sheath and ICME boundary, represented by vertical lines. 

At around 09:53 UT on 13 March 2022, the shock arrived at L1 with a speed of 586 km/s, indicated by the sudden jump in magnetic field magnitude and other solar wind parameters. The shock is followed by a period of magnetic field magnitude fluctuations, and all monitored solar wind parameters, such as proton plasma $\beta$ parameter below 1, signifying the sheath region. Subsequently, starting from 13 March 22:35 UT, the sheath transitioned into the ICME region with speeds reaching up to 450 km/s by 13 March 23:00 UT. Immediately after the ICME arrival, that is at around 13 March 22:35 UT; there was a significant rotation in the magnetic field components $B_z$ and $B_y$ ($B_z$ shifted from northward to southward and $B_y$ southward to northward). This rotation of the magnetic field components suggested the possible presence of an MC within the ICME structure. During this rotation, proton density, temperature, and proton-$\beta$ plasma parameter exhibited abrupt increases, while the latter remained below 1. On 14 March at about 06:47 UT, the magnetic field magnitude attained a maximum  $B_{0_{WIND}}=$ 24.41 $\pm$ 0.05 nT.

The uncertainty shown is the standard uncertainty associated with MFI/WIND \citep{WINDMAGNETICFIELD} measurements. Following the rotation phase, the magnetic field evolved more smoothly. The ICME event concluded at 23:46 UT on March 15 when the WIND-measured magnetic field
transitioned into an area with consistently stable lower magnetic field values, and the plasma $\beta$ value exceeded 1. 


As previously mentioned, a recent study by \citet{Laker2023} examined the same event. They employed a magnetic cloud fitting using the Lundquist model to analyse in situ magnetic field data from both SolO (0.43 AU) and WIND (0.99 AU). They found impact parameter values, which indicate how close the spacecraft came to the central axis of the flux rope, of -0.18 $\pm$ 0.02 R$_E$ (where R$_E$ represents Earth's radius) for SolO and -6.3 $\pm$ 0.24 R$_E$ for WIND. \citet{Zhuang2024} estimated the average radial size of this ICME as 0.125 AU ( $3.05 \times 10 ^{3} R_E$) and 0.35 AU ( $8.2 \times 10^{3}R_E$) at SolO and WIND, respectively. Considering the radial size of the ICME at both SolO and WIND, the impact parameter is significantly smaller, indicating a near-axis crossing. Interestingly, the orientation of the flux ropes was consistent between the spacecraft, indicating a stable magnetic structure. These findings underscore that despite a separation of 0.5 AU, the magnetic structure of the flux rope remained relatively stable across the spacecraft, with the observed maximum magnetic field strength essentially corresponding to the axis of this structure. Therefore, our approach of considering the maximum magnetic field from in situ measurements as axial fields, which is later compared with the axial field from the theoretical flux rope model (Lundquist) ( see Sect.\ref{Results -estimation of near sun B}), is justified.


Lastly, we mention that at the time of this CME, BepiColombo was about 5$^o$ east of the Sun-STEREO A, which was 34$^o$  east of Earth, and 6.8$^o$ north of the STEREO A ecliptic latitude. (see Fig \ref{spacecraftposiion}). The recent study of \citet{Jackson2023} has explored the in situ magnetic field and electron density measurements of the event from BepiColombo data. However, from the orientation of the CME (latitude, longitude, tilt), the angular width ( detailed in Sect.\ref{Results-forward modelling}) and the position of the spacecraft, we infer that the possible arrival of our analysed ICME at BepiColombo could have resulted in a flank impact from its western side. Therefore, BepiColombo measurements may not be suitable for estimating the axial magnetic field. For this reason, we have not used BepiColombo data in our analysis. 

We would now like to mention in situ measurement indications of the possibility of a complex ICME structure. Distinctive features in the ICME sheath structure are observed in WIND data (see Fig.\ref{WINDinsitu}), which includes prolonged sheath and magnetic ejecta that exceed typical lengths beyond the end of the rotation period, suggesting a more complex structure. Analysing temperature trends during the ICME reveals a monotonous increase in the start, followed by steep changes. Simultaneously, the velocity profile during the expansion phase exhibits significant complexity, hinting at a possible composite ICME structure. These intricate temperature and velocity patterns further support the notion of a dynamically evolving CME during propagation. These interpretations are not further tested/analysed extensively as they are beyond the scope of this study.

\section{Methodology}\label{Methodology}
Throughout this work, we follow the methodology introduced by \citet{Patsourakos_2016} to estimate the near-Sun axial magnetic field of the CME based on the helicity conservation principle. This session is divided into two subsections. In Sect.\ref{Methodology- Near sun analysis}, Sect. \ref{Method- magnetichelicty} outlines the calculation of the magnetic helicity ($H_m$) in the CME source region, followed by an assessment of the helicity budget difference between the pre-and post-eruption phases in the source AR. Sect.\ref{method-forward modelling} describes the determination of the CME's geometrical characteristics by applying the GCS model fitting to multi-viewpoint coronagraph observations to derive parameters such as flux-rope radius and length. Sect.\ref{Methods -estimation of near sun B}, which explains the methodology for estimating the near-Sun axial magnetic field at the maximum height obtained from forward modelling of the CME using the Lundquist flux rope model \citep{lundquist1950_fluxrope}. In Sect.\ref{Methods- extrapolation}, the magnetic field extrapolation method is described. This method is used to understand the magnetic field evolution during CME propagation to two vantage points, namely, 0.43 AU and 0.99 AU. It assumes a power-law variation of the magnetic field with the heliocentric distance, utilising in situ magnetic field measurements obtained from MAG/SolO and MFI/WIND.

\subsection{ Near -Sun analysis}\label{Methodology- Near sun analysis}
\subsubsection{Magnetic helicity budget of the CME}\label{Method- magnetichelicty}

To understand the temporal evolution of relative magnetic helicity and free magnetic energy budgets of AR 12962, we use the CB discrete flux tube method developed by \citet{Georgoulis_2012}. This CB method computes the instantaneous relative magnetic helicity and free magnetic energy budgets above a potential field reference by using a single vector magnetogram, in which the magnetic flux distribution is partitioned. This method generates a connectivity matrix, accounting for magnetic flux connections between positive and negative polarity partitions. This computation employs a simulated annealing method to prioritise connections between opposite polarity partitions while minimising connection lengths. The resulting connections form an ensemble of N (assumed slender) force-free flux tubes, each characterised by known footpoint locations, magnetic flux, and force-free parameters. The relative magnetic helicity, $H_m$ is the sum of self and mutual terms, corresponding to the twist of the flux tube and linkage between different tubes, respectively. Since helicity is a signed quantity, the negative sign in $H_m$ corresponds to a left-handed helicity, and the positive sign corresponds to a right-handed helicity. Hereafter, we use the term 'helicity' to refer to the relative magnetic helicity.

\begin{figure*}[ht]
\centering
  \includegraphics[width=\textwidth]{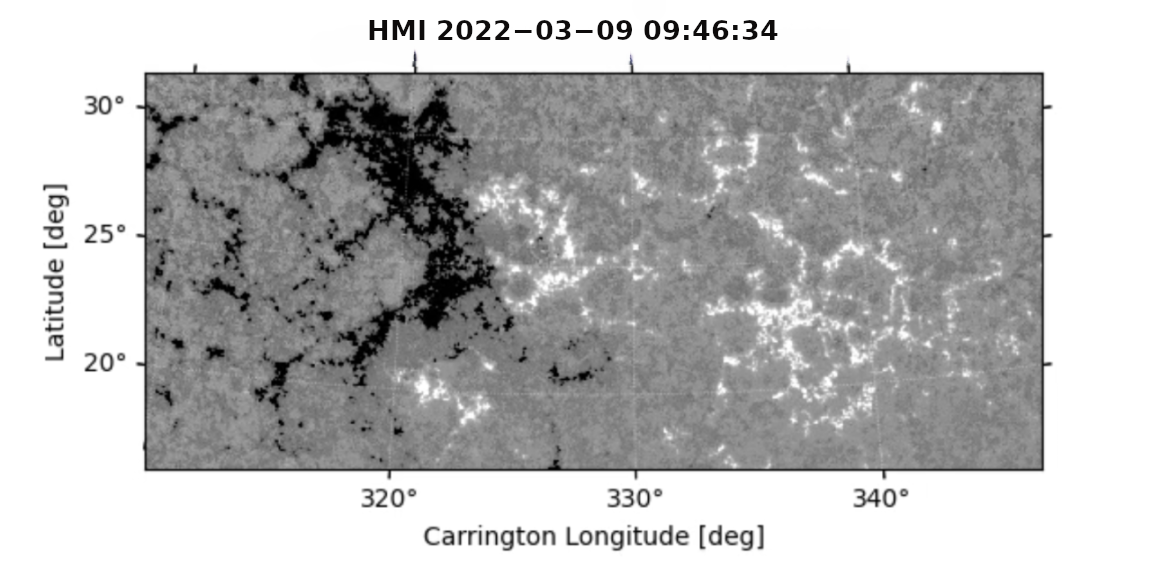}
 \caption{$B_{z}$ component of photospheric vector magnetogram of AR 12962 on 2022-03-09 09:46 UT.
}
\label{hmi}
\end{figure*}

To calculate the energy and helicity budgets, we used SDO/HMI vector magnetograms \citep{Hoeksema14}, specifically the HMI.SHARP\_CEA\_720s data series \citep{Bobra14}. This dataset includes Lambert cylindrical equal-area (CEA) projections of the photospheric magnetic field vector. This HMI data product has been transformed into three spherical heliographic components, $B_r$, $B_{\theta}$, and $B_{\phi}$. These components are related to the heliographic field components as [$B_x$, $B_y$, $B_z$] $=$ [$B_{\phi}$, $B_{\theta}$, $B_r$], where $x$, $y$, and $z$ indicate the solar westward, northward, and vertical directions, respectively \citep[see][]{Sun13}. The spatial resolution of the CEA vector field images is 0.03 degrees, equivalent to about 360 km per pixel at the disk centre, and the cadence of the magnetogram time series is 12 min. Fig. \ref{hmi} represents the $B_z$ component of the AR 12962 on 09 March 2022 at 09:36:34 UT. The AR is seen to be bipolar, with a dispersed magnetic field distribution, and the maximum vertical magnetic field in the shown magnetogram is 1242.01 Gauss.

For this study, we used 134 magnetograms from 10 March 2022, 08:00UT to 11 March 2022, 10:46UT. This includes 41 magnetograms before the dimming start time, 13 magnetograms between the dimming start and peak time, and 82 magnetograms after the dimming peak time. To implement the CB method, we used specific thresholds for magnetogram partitioning to streamline the analysis and reduce computation time: (1) a threshold of 50 G for the minimum $B_z$ participating in the partition map, (2) a minimum magnetic flux of $2\times10^{19}$ Mx per partition, and (3) a minimum area of 30 pixels per partition, that is $\sim 3.9 \times 10^6\;\mathrm{km^2}$ in solar surface, and (4) typical uncertainties for vertical and horizontal fields set at 5 G and 50 G, respectively \citet{Georgoulis_2012}. These thresholds aim to exclude quiet-Sun, weak-field regions, and very small-scale structures from the calculation. The thresholds are optimised to include the majority of the total unsigned flux of the vector magnetogram while maintaining computational efficiency. Through a trial and error method, thresholds were identified beyond which additional regions do not significantly affect helicity and energy values, which remain relatively stable. For example, changing the minimum flux per partition threshold from $2\times10^{19}$ Mx to  $1\times10^{19}$ Mx increases the helicity value by an average of 15-20\%, while the ratio of partitioned flux to total unsigned flux rises from $\sim$70\% to $\sim$85\%. This increase in helicity is rather insignificant compared to the increase in computational time due to the inclusion of the quiet-Sun regions. Hence, we fixed our minimum flux per partition as $2\times10^{19}$ Mx. We calculated the standard deviations of the moving five-point averages (48 min) to determine uncertainties for all quantities. This approach is preferred over the default uncertainty mentioned in \citet{Georgoulis_2012},  which tends to give rise to smaller uncertainty amplitudes.

\subsubsection{Forward modelling of the CME}\label{method-forward modelling}
In order to determine the geometrical parameters of the CME,  we used the GCS model of \citet{Thernisien_2006} and \citet{ Thernisien2009}. This geometrical representation of flux rope CME fits the CME envelope with simultaneous observations from two or three vantage points. The model relies on various free geometrical and positional parameters. The positional parameters determine the location and orientation of the ﬂux rope, which is achieved by supplying the model flux rope's apex's heliographic (Stonyhurst) longitude $\phi$ and latitude $\theta$ and the orientation of its axis of symmetry (tilt) $\gamma$. The geometrical parameters such as the height of the flux rope ($h$), the angular width between its two legs ($2w$), and its aspect ratio ($\kappa$), control how fast the structure expands relative to its height, assuming a self-similar expansion. The user varies these six parameters on a trial-and-error basis until a good agreement is reached between the multi-viewpoint observations and the corresponding model projections.
From model-fitted results, we calculate the radius $R$ and length $L_{CME}$ of the CME flux rope. The radius $R$ is calculated using the relation,
\begin{equation}
R = \frac{h}{1 + \frac{1}{\kappa}}
\label{R0 eqn}
\end{equation} which is derived using Equation (1) of  \citet{Thernisien_2006}. 

We calculated the length of the CME flux rope using two approaches: 1) as in \citet{Patsourakos_2016} and 2) as in \citet{Pal_2017}. The \citet{Patsourakos_2016} approach assumes that the  CME front is a cylindrical section with an angular width given by the GCS ﬁtting. The length of the CME flux rope  according to \citet{Patsourakos_2016} is given by:

\begin{equation}
    L_{CME}=2 tan (w) r_{mid}
    \label{pat L eqn}
\end{equation}
where $r_{mid}= h - R$ is the heliocentric distance halfway through the model’s cross-section along its axis of symmetry, and $w$ is the half angular width in radians. On the other hand, using the approach in  \cite{Pal_2017} we have, 
\begin{equation}
L_{\mathrm{CME}}=2 h_{\mathrm{leg}}+y\left(h-h_{\mathrm{leg}} / \cos \gamma\right) / 2-2 R_{\odot}
\label{pal L eqn}
\end{equation}
where $h_{leg}$ is the height of the legs of the CME ﬂux rope computed using Equation (3) of \cite{Thernisien_2006},
$\left(h-h_{\mathrm{leg}} / \cos \gamma\right) / 2$ is the radius of the arc of the ﬂux rope, $y=2(\pi / 2+\gamma)$ is the arc angle in radians, and $R_{\odot}$ represents the solar radius.

To introduce the magnetic field of the flux rope, we consider the Lundquist cylindrical model 
\citep{lundquist1950_fluxrope}. This is an axisymmetric linear force-free solution with the following form in cylindrical coordinates (r, $\phi$, z):
\begin{equation}
B_r=0, B_\phi=\sigma_H B_0 J_1(\alpha r), B_z=B_0 J_0(\alpha r)
\end{equation}
where $B_0$ is the maximum axial ﬁeld, $J_0$ and $J_1$ are the Bessel functions
of the zeroth and ﬁrst kinds, respectively, $\sigma_H=\pm 1$ is the helicity
sign, and $\alpha$ is the constant force-free parameter. Assuming that the ﬁrst zero of $J_0$ is reached
at the edge of the ﬂux rope, namely
\begin{equation}
    \alpha R = 2.405
    \label{alphaR}
\end{equation}
then $R$ corresponds to the ﬂux-rope radius. This essentially leads to a purely axial (azimuthal) magnetic ﬁeld at the ﬂux-rope axis (edge).

The Lundquist flux rope model is widely used to describe flux ropes in ICMEs. \citet{PatsourakosPrametric} analysed different analytical flux rope models for the same input parameters to estimate the axial magnetic field of modelled CMEs. They found that the Lundquist flux rope model gave the highest axial magnetic field values among the other models. Recently, \citet{LYNCH2022} supported its application in the near-Sun environment: their study 
evaluates the performance of the Lundquist flux rope model by fitting it to synthetic spacecraft trajectories derived from the spatial coordinates of PSP's encounters. The results of the study indicate that the in situ flux rope models, including the Lundquist model, are generally a decent approximation to the magnetic field structure within 30 $R_{\odot}$ compared to MHD models.  

\subsubsection{Estimation of Near-Sun magnetic field}\label{Methods -estimation of near sun B}



Assuming the Lundquist flux rope 
model with  $L_{CME}$, $R$ derived from GCS fitting, to obtain the magnetic helicity of the cylindrical flux rope, we follow
\citet{ DeVore_2000}, \citet{Demoulin_2002} and \citet{Dasso_2003} as
\begin{equation}
    B_{0} = \sqrt{\frac{H_m}{0.7 \cdot R^3 \cdot L_{CME}}} .
    \label{eqnof B estimtion}
\end{equation}

A Monte Carlo simulation determines the associated uncertainty in the estimated CME axial magnetic field $B_0$. This approach generates multiple samples by randomly selecting input parameter values from assumed uniform probability distributions of input GCS parameters. The procedure is described in detail in Appendix \ref{monte-carlo procedure}.


\subsection{Extrapolating the CME axial field to 0.43 and 0.99 AU
}\label{Methods- extrapolation}


As CMEs propagate in the heliosphere, their magnetic field decreases with increasing distance from the Sun and subsequent expansion. A number of previous studies such as 
\citet{BothmerSchenn1998}, \citet{Wang2005}, \citet{Farrugia2005ESASP}, \citet{demoulinDasso2009}, \citet{winslow2015} and \citet{PatsourakosPrametric} find that this decrease resembles a self-similar behaviour, that is, it can be approximated by a power law. Therefore, we extrapolate the estimated axial magnetic field of the CME at GCS fitted height 0.03AU to 0.43 AU and 0.99 AU by assuming a power-law variation of the axial magnetic field with the heliocentric distance r given by

\begin{equation}
B_0(r)=B_*\left(r / r_*\right)^{\alpha_B}.
\label{extrpolationEqn}
\end{equation}

Here, $B_*$ corresponds to the estimated axial magnetic field at GCS fitted height r$_*$( from Eq.\ref{extrpolationEqn}), and B$_0(r)$ corresponds to the maximum ICME magnetic field from in situ measurements at distances (r) of 0.43 AU and 0.99 AU, respectively. $\alpha_B$ is the power-law index.

In this extrapolation, we utilised the approach described in \cite{Patsourakos_2016} to find optimal values for $\alpha_B$. We examined $\alpha_B$ within the range of [-2.7, -1.0], with a step size of 0.1. This specific range was derived primarily from diverse theoretical and observational studies, encompassing both the outer corona and the inner heliosphere and, in some cases, addressing both simultaneously \citep[][]{Patzold1987SoPh, KumarandRrust1996, Bothmer1997TheSA,Vsnak2004A&A, Liu2006JGRA, Forsyth2006ICMEsIT, Leitner2007, Poomvises2012DETERMINATIONOT, Mancuso2013CoronalMF}. In addition, we employed linear regression to the log-log pairs to three points in the inner heliosphere, namely, the estimated $B_0$ at GCS fitted height, ICME $B_{0_{SolO}}$ at 0.43 AU from MAG/SolO, and ICME $B_{0 WIND}$ at 0.99 AU from WIND.

\section{Results}\label{Results}

\subsection{ Near-Sun analysis}\label{Results- Near sun analysis}
\subsubsection{Magnetic helicIty budget of the CME}\label{Results- magnetichelicty}

To comprehend the pre- and post-eruption behaviour of the source AR, we pinpointed temporal milestones of the eruption, such as the CME onset. The time stamps of coronal dimmings can serve as a proxy indicator for CME onset. Coronal dimmings are a widely recognised indicator of the initiation of CMEs \citep{Howard2004}, which refers to areas with significantly decreased emission in EUV wavelengths \citep{Thompson1998} and soft X-rays  \citep{Hudson1996ApJ, Sterling1997}. These dimmings, observed in the corona, occur in conjunction with CMEs and serve as indicators of the extraction and depletion of coronal plasma due to the CME expansion and ascent in the corona \citep{Thompson1998, Harrison2000A&A}. Hence, we consider the dimming start and peak time as the eruption onset and maximum, respectively, which is then used to define the pre- and post-eruption phase of the source AR. We used the Solar Demon Dimming Detection tool \footnote{\url{https://www.sidc.be/solardemon/dimmings.php}} \citep{Solardemondimming2015JSWSC} which runs in real-time on SDO/AIA 211 \AA~data with 3-minute cadence to obtain the intensity of the darkest region within the dimming regions in AR 12962 (see Fig.\ref{CME remote sensing 3 images}b). 
The  GOES 1 – 8 Å X-ray flux X-ray sensor (XRS) 1-minute data shows the associated flare timing. \footnote {\url{https://www.swpc.noaa.gov/products/goes-x-ray-flux}}


\begin{figure*}[h!t]
\centerline{
\includegraphics[width=17cm]{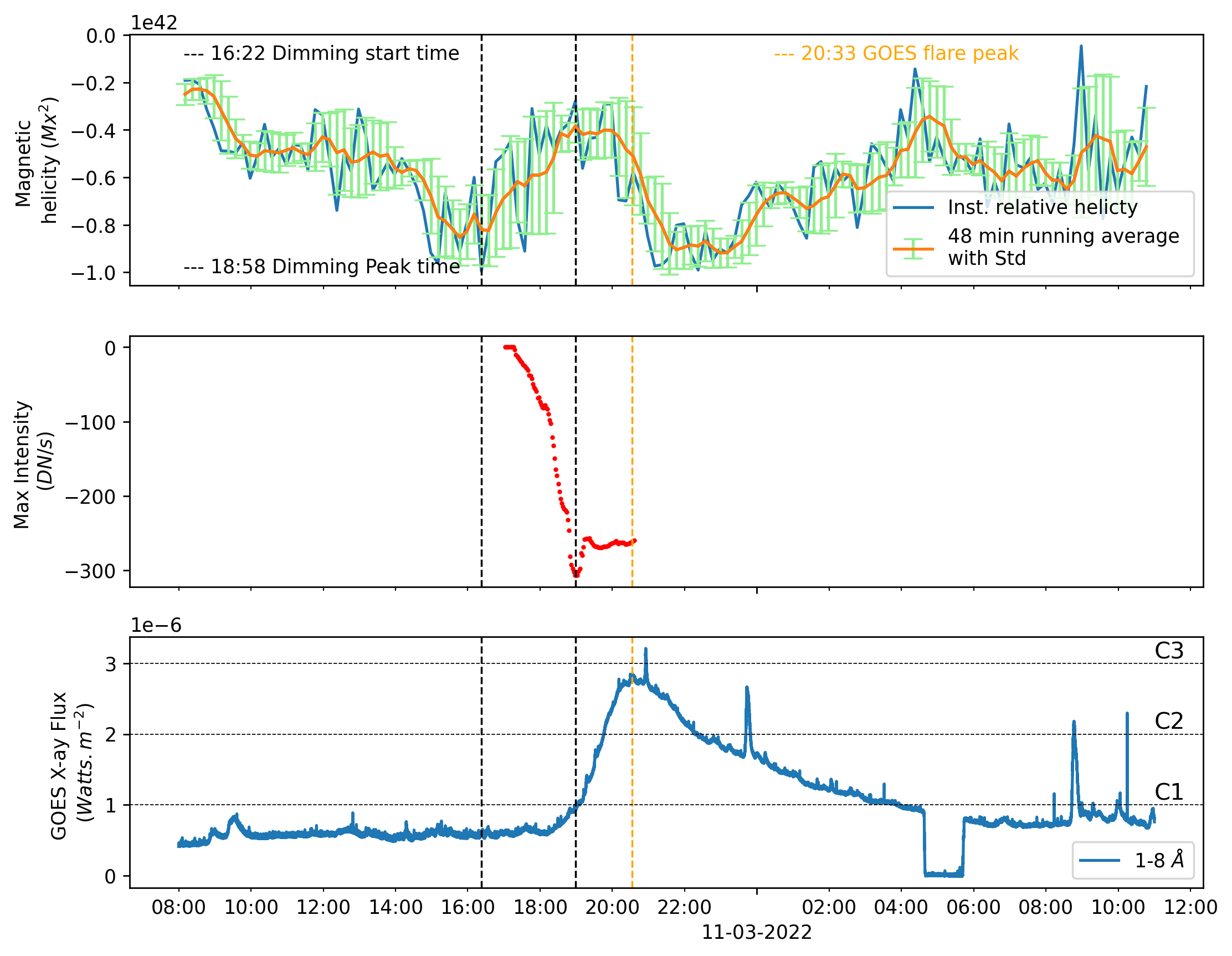}
}
\caption{Temporal evolution of magnetic properties of NOAA AR 12962. Top to bottom: a) Blue lines represent the instantaneous values of the relative magnetic helicity estimated from the CB method. The over-plotted orange curve and green error bars are the running averages of the estimated values over a 48-minute window, with the standard deviation of each 48-minute moving average, respectively.
b) Red dots represent the base difference intensity of the darkest region within the dimming.
c) The blue curve represents GOES 1 – 8 \AA~X-ray flux 1-minute data. In all panels, vertically dashed black lines indicate the dimming start and peak time, and the orange line represents the GOES flare peak time.}
\label{helicty plot 1}
\end{figure*}

Fig.\ref{helicty plot 1}a represents the temporal evolution of instantaneous values of relative magnetic helicity of AR 12962 from 11 March 2022 08:00 UT to 11 March 2022 11:00 UT, along with a 48-minute moving average and respective uncertainty. Following this, Fig.\ref{helicty plot 1}b and Fig.\ref{helicty plot 1}c
illustrate the temporal progression of the base difference intensity of the darkest region within the dimming and the GOES X-ray flux, respectively. The dimming start and peak times are marked as vertical black dashed lines in all panels.

An inspection of Fig.\ref{helicty plot 1}a shows that the net helicity sign is negative throughout the time period under investigation. Since the AR is in the northern hemisphere, this is consistent with the hemispheric helicity rule (\cite{Pevtsov2014MagneticHT} and references therein). The helicity shows that it is decreasing in magnitude starting on 10 March 2022, at 15:10 UT until 18:58 UT  the same day. During this period, the absolute value of net helicity continuously decreased from $(-0.99 \pm 0.12) \times  10^{42} \;\mathrm{Mx^{2}}$ to $(-0.28 \pm 0.12) \times10^{42}\; \mathrm{Mx^{2}}$, so the net helicity decrease is $(-0.71 \pm 0.12) \times  10^{42}\; \mathrm{Mx^{2}}$. The uncertainty in this value is calculated by employing error propagation in the pre-and post-eruption helicity values. The interval of $H_m$ decrease is roughly bracketed between the timestamps of start and peak dimming intensities (check out  Fig. \ref{helicty plot 1}b), which 
implies a potentially close connection between the CME onset and early evolution of $H_m$. We note that the difference between the pre-and post-eruption helicity is well above the applicable uncertainty. As evidenced from full-disk images in the EUV, transfer of this helicity to global solar structures via magnetic reconnection is rather limited as the unstable structure seems fairly isolated. Hence, we assume that the difference in helicity is entirely transferred to the CME. It is noted that after the dimming's peak time, at around 18:58 UT, total helicity enters a recovery phase of 2 h 48 min and reaches a maximum close to that of the pre-eruption stage, at $(-0.99 \pm 0.12) \times  10^{42}\; \mathrm{Mx^{2}}$ around 21:46 UT. 
From Fig.\ref{helicty plot 1}c, which shows the GOES X-ray flux, we infer that the XRS values took several hours past the flare peak to return to pre-flare values, inferring this was a long-duration event typical of eruptive flares.

\begin{figure*}[h!t]
\centerline{
\includegraphics[width=17cm]{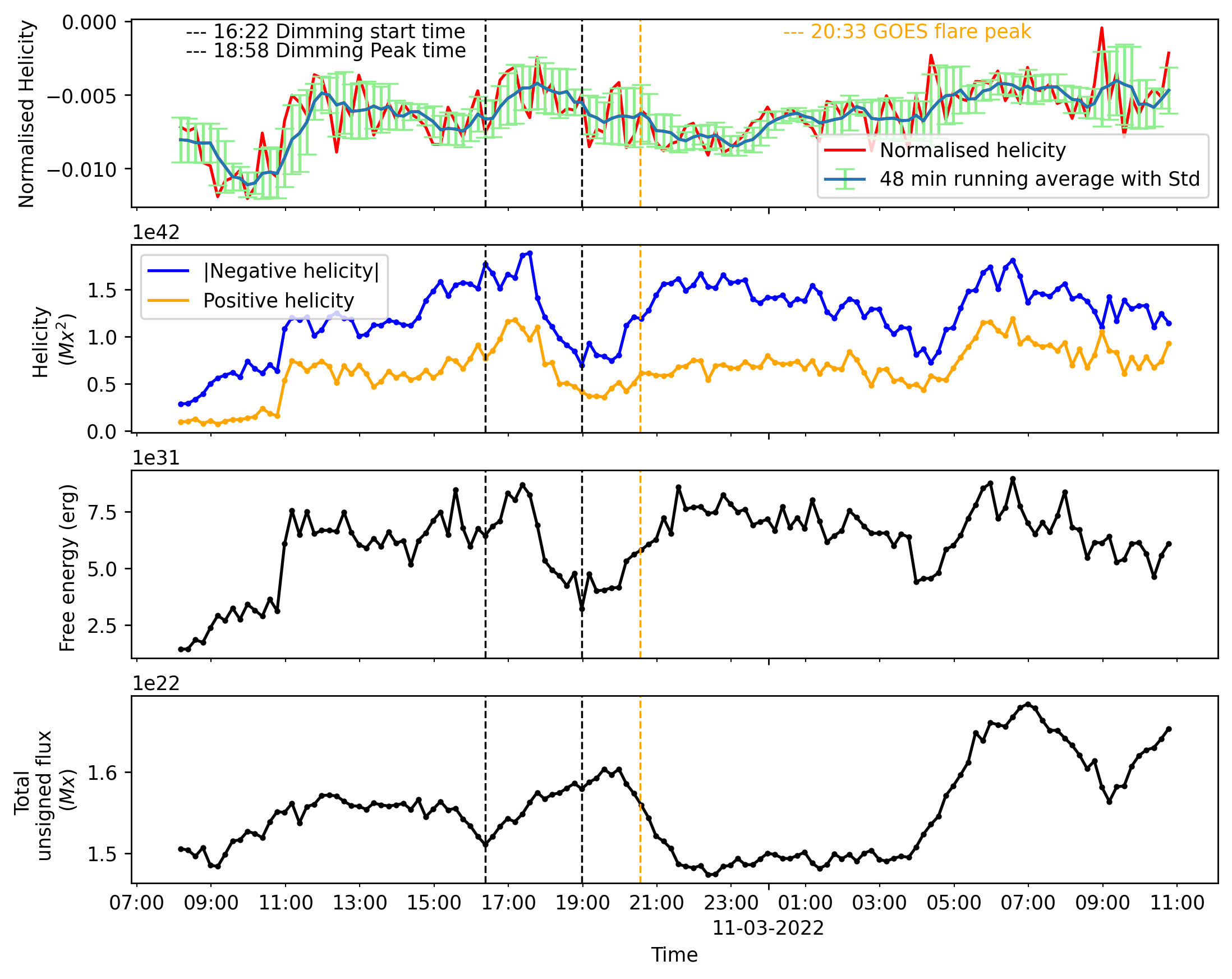}
}
\caption{Temporal evolution of other magnetic attributes of NOAA AR 12962. Top to bottom:
a) The red curve is the normalized helicity value. The blue overplotted curve and the green error bar correspond to the  48-minute moving average and the standard deviation, respectively. b) The blue line represents instantaneous normalised negative helicity, and the orange line represents the normalized positive helicity. c) The black line represents the normalized instantaneous magnetic free energy. d) The black line represents the instantaneous total unsigned flux.
In all panels, the vertically dashed black lines indicate the dimming start time (16:22) and dimming peak time (18:58), and the dashed orange line represents the GOES flare peak time.}
\label{helicty plot2}
\end{figure*}

Fig.\ref{helicty plot2} illustrates the temporal variation of other magnetic properties of the AR. Fig.\ref{helicty plot2}a displays the normalised helicity variation, calculated as the ratio of the instantaneous relative magnetic helicity to the square of the total connected flux. The normalised helicity provides a generalised helicity measure independent of AR flux distribution. As seen in  Fig.\ref{helicty plot2}a, this time series is more featureless (that is, less responsive to eruptions) than the actual helicity budget of Fig.\ref{helicty plot 1}a, as also suggested by \citet{LaBonteGeorgoulisRust2007} and \citet{Thalmann19}.

In Fig.\ref{helicty plot2}b, the components contributing to the total helicity are presented separately, namely, the total negative helicity and total positive helicity. Throughout the observed period, the dominant helicity is negative (left-handed); there is a declining trend in left- and right-handed helicity before the eruption, resulting in a decreasing net helicity, with left-handed helicity (negative sign) domination. Fig.\ref{helicty plot2}c represents the temporal evolution of free magnetic energy, starting from 10 March 2022, at 16:58 UT, which has decreased from  $ 8.33 \times 10^{31} \mathrm{erg}$ to $4.15 \times 10^{31} \mathrm{erg}$  at 19:03 UT, marking a difference between pre-and post-eruption free energy budget as  $4.18 \times 10^{31} \mathrm{erg}$ which is within the range of CME mechanical energies discussed in \citet{Vourlidas2010energycme}. 
Similarly to the $H_m$, the period of free magnetic energy decrease is co-temporal with the period of dimming start-peak. It then enters a stage of recovery to a maximum of $8.6 \times 10^{31} \mathrm{erg}$ at 21:34 UT. 

Fig.\ref{helicty plot2}d shows the total unsigned flux variation. It initially goes through an increasing phase just before the eruption. It has a minimum value of  $1.51\times 10^{22}\; \mathrm{Mx}$  on 10 March 16:22 UT and reaches a maximum of $1.6 \times10^{22}\; \mathrm{Mx}$ on 10 March 19:58 UT. After the eruption, flux decreases to approximately its initial value and persists in that value for nearly six hours. This is likely attributed to the enhanced horizontal magnetic field in the post-eruption AR caused by magnetic implosion \citep{Hudson_2000}. As flux entirely depends on the vertical component of the magnetic field ($B_z$), a smaller flux may indicate a reduced $B_z$ after the eruption. This can also explain the absolute helicity value increase around the flare peak, which continued after the CME, potentially due to an enhanced horizontal field and increased force-free parameters right after the eruption, as seen by \citet{Liu2023}. This can be attributed to the delayed interaction between the photosphere's ongoing processes and the coronal events.

\subsubsection{Forward modelling of the CME}\label{Results-forward modelling}
To estimate the geometrical parameters of the CME, we used white light observations from SECCHI/COR2 and LASCO/C2 coronographs, which have a FOV of 2.5-15  $R_{\odot}$ and 2-6 $R_{\odot}$, respectively. STEREO-A was $33.0^{\circ}$ ahead of Earth at the time of the CME. The CME was observed nearly simultaneously as a west limb event in STEREO-A and as a partial halo CME in the SOHO C2 and C3 coronagraphs. However, in the LASCO C3 observations, the CME appeared more diffuse, making its boundaries harder to identify reliably and, therefore, unsuitable for a GCS fit.

\begin{figure*}[h!t]
  \centering

  \begin{subfigure}{0.48\textwidth}
    \centering
    \includegraphics[width=\linewidth]{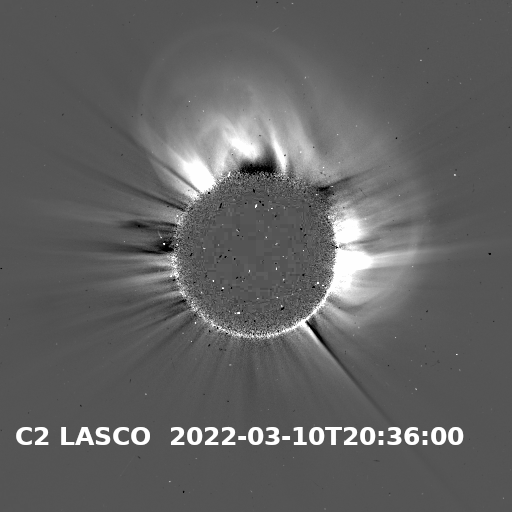}
    \label{lascogcs}
  \end{subfigure}
  \hfill
  \begin{subfigure}{0.48\textwidth}
    \centering
    \includegraphics[width=\linewidth]{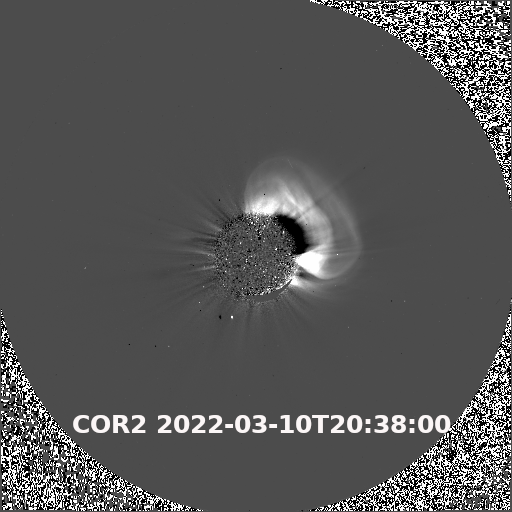}
    \label{strgcs}
  \end{subfigure}

  \vspace{1em}

  \begin{subfigure}{0.48\textwidth}
    \centering
    \includegraphics[width=\linewidth]{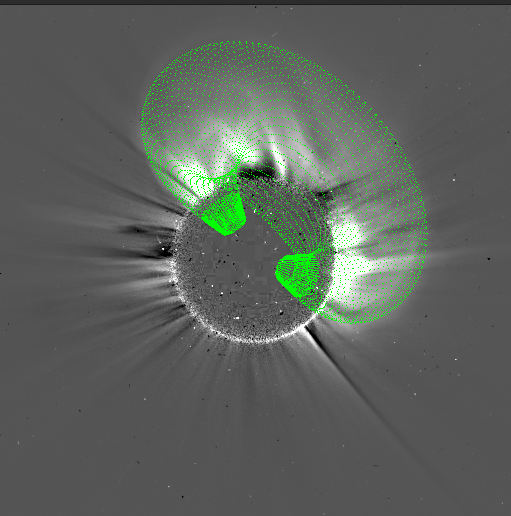}
    \label{gcslasco}
  \end{subfigure}
  \hfill
  \begin{subfigure}{0.48\textwidth}
    \centering
    \includegraphics[width=\linewidth]{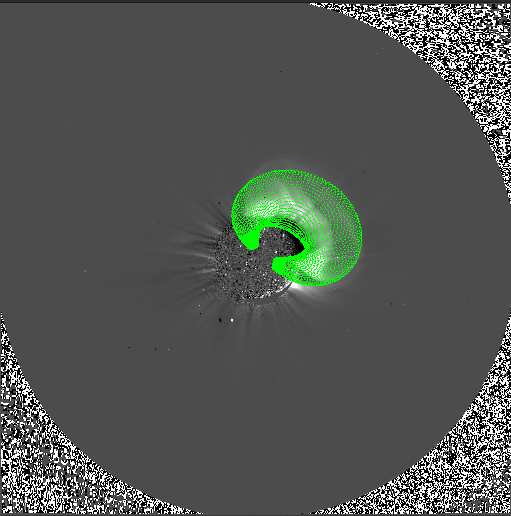}
    \label{gcsstr}
  \end{subfigure}

  \caption{GCS model fits on the 10  March 2022 CME. Top row: Images of the CME by LASCO/C2 (left) and SECCHI COR2/STEREO-A (right). Bottom row: 
  The same images, with the GCS wireframe, overlaid.}
  \label{GCSFitted}
\end{figure*}

In Fig.\ref{GCSFitted}, the top row represents two nearly simultaneous observations of the CME, one with 
LASCO C2 at 20:36 (left) and another with STEREO-A at 20:38 UT (right). The bottom row includes the overplotted GCS wire frame. 
The best-fit parameter values and applicable uncertainties are listed in Table \ref{GCS params table}. 

The uncertainties are estimated manually, similar to the sensitivity analysis outlined in \citet{Thernisien2009}. Specifically, our approach involves perturbing the optimal fit of one parameter at a time, up to an extent the deviations are within visually acceptable range while keeping the other parameters constant at their optimal fit. The uncertainties in the parameters are hence measured in terms of deviations from their optimal values, considering both positive and negative extremes. The fitted results of the position, tilt, and angular width of the CME are consistent with the conclusion that this CME is propagating towards SolO. Our obtained GCS fitting results are also consistent with those of \citet{Zhuang2024}.

\begin{table}
\caption{CME GCS fitted parameters along with the positive and negative limits of uncertainty.}             
\label{GCS params table}      
\centering                                      
\begin{tabular}{c c c }          
\hline\hline                        
    \textbf{CME parameter} & \textbf{Value}  \\
    \hline
    latitude (degrees) & 20.7  $\pm$4.3 \\ 
 
    longitude (degrees) & 7.8  $\pm$ 1.8\\ 

    tilt angle (degrees) & -50.9  $\pm$22 \\
    height (solar radii) & 7.6  $\pm$ 0.4 \\
    aspect ratio & 0.3, $-0.05$, $+0.03$ \\
    half-angle (degrees) & 55.6 , $-10.62$, $+9.22$ \\
    \hline
    R (solar radii) & 1.8  $\pm$ 0.4 \\
    L\textsubscript{CME} (Solar radii) & 17.0  $\pm$ 3.7 \\
    \hline

\hline                                             
\end{tabular}
\end{table}

From GCS fitting results, we obtain the height of the CME as 7.6 $R_{\odot}$, that is, at a distance of 6.6 $R_{\odot}$from the solar surface. We calculate the radius $R$ using Eq.\ref{R0 eqn}. The length $L_{CME}$ is calculated using two approaches mentioned in Eq.\ref{pat L eqn} and Eq.\ref{pal L eqn}. We employed a Monte Carlo simulation to estimate uncertainties in these derived parameters. The variations in these parameters are simulated by assuming a uniform distribution within specified ranges. We obtained the length of the CME  following \citet{Patsourakos_2016} as 17.02 $\pm$ 3.85 $R_{\odot}$ and by following \citet{Pal_2017} we obtained 13.65 $\pm$ 0.68 $R_{\odot}$. 

\subsubsection{Estimation of the near-Sun magnetic field of the CME} \label{Results -estimation of near sun B}
We now have three approaches for estimating the near-Sun axial magnetic field of the CME from Eq. \ref{eqnof B estimtion}. First, we estimate the helicity budget of the CME by assuming the net helicity difference between the pre- and post-eruption phase of AR, yielding $(-7.1 \pm 1.2) \times 10^{41} \mathrm{Mx^{2}}$. Under this assumption, the axial magnetic field at 6.6 $R_{\odot}$ is determined as $B_0 = 2067 \pm 405$ nT, with the CME length calculated following Eq. \ref{pat L eqn}. Secondly, adopting the same helicity budget assumption but employing the length of the CME derived from Eq. \ref{pal L eqn}, we obtain $B_0 = 2600 \pm 440$ nT. Thirdly, we estimate the magnetic field value by attributing the total pre-eruption helicity budget to the CME, as proposed by \citet{Patsourakos_2016}, resulting in $B_0 = 2437 \pm 381$ nT, while still following the length calculation method from Eq. \ref{pat L eqn}. Both methods for calculating the length (first and second) and the third method of assigning pre-eruption helicity to the AR yield similar $B_0$ values within the uncertainty range. Therefore, we choose the second method, that is $B_0 = 2067 \pm 405$ nT.

\subsection{Extrapolating near-Sun CME magnetic field to 0.43 AU and 0.99 AU}\label{Results- extrapolation}

\begin{figure}[h!t]
\centerline{
\includegraphics[width=0.5\textwidth]{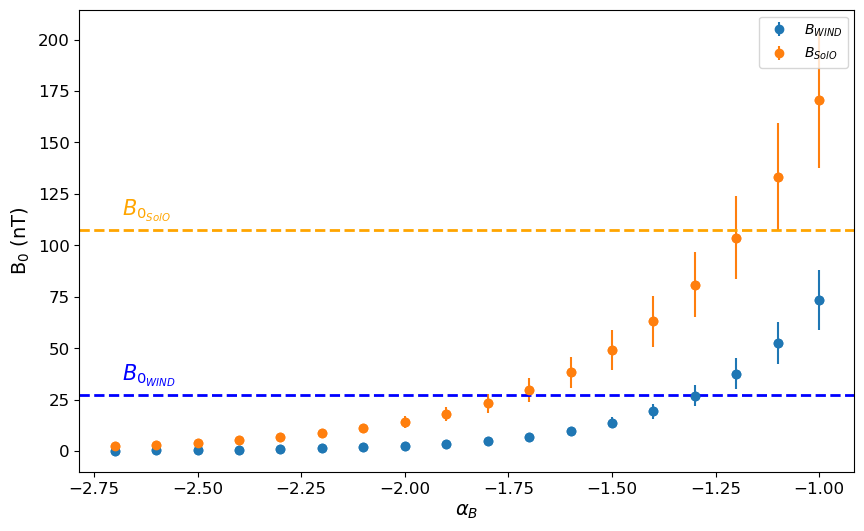} }

\caption{Extrapolated CME magnetic ﬁeld at 0.43 AU (orange points) and 0.99 AU (blue points) as a function of the radial power-law index ($\alpha_B$) of Eq.\ref{extrpolationEqn}. The orange and blue horizontal dashed lines correspond to the ICME maximum magnetic field measurement from SolO and WIND, respectively.}
\label{B0rPlotExtrapolation}
\end{figure}

As explained in Sect.\ref{Methods- extrapolation},  using Eq.\ref{extrpolationEqn}, we plotted the extrapolated CME magnetic field as a function of the power-law index $\alpha_B$ in Fig.\ref{B0rPlotExtrapolation}. The figure marks these maximum magnetic fields as dashed lines. From Fig.\ref{B0rPlotExtrapolation}, the $\alpha_B$ value that intersects with the maximum magnetic-field measurement of SolO is -1.1, and that of WIND is -1.2. 

\begin{figure}[h!t]
\centerline{
\includegraphics[width=0.5\textwidth]{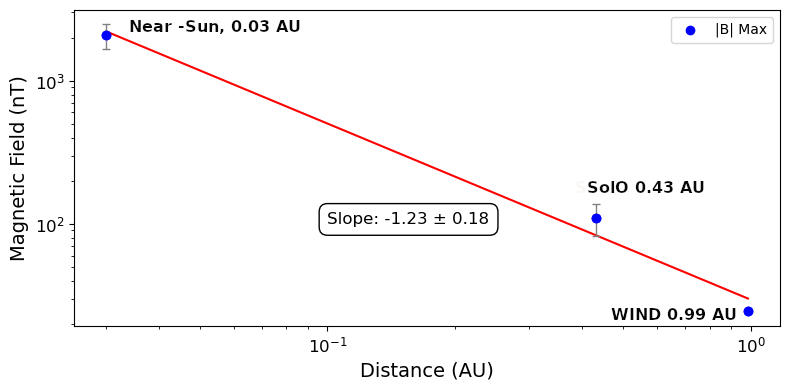}
}
\caption{CME magnetic field variation with distance. Blue points with error bars depict the maximum magnetic field $B_0$. The near-Sun field is estimated, while the fields at 0.43 AU (Solar Orbiter) and 0.99 AU (WIND) rely on in situ measurements. The 0.99 AU measurements have minimal error (on the order pT), and the red line represents the least-squares best fit.}  
\label{B0rPlot}
\end{figure}

In addition, we employed linear regression to the log-log pairs of the maximum magnetic field measurements from three points in the inner heliosphere, namely, the estimated near-Sun axial magnetic field at GCS fitted distance of 6.6 $R_{\odot}$ or 0.03 AU ($B_0$), $B_{0 SolO}$ at 0.43 AU, and $B_{0 WIND}$ at 0.99 AU. This is illustrated in Fig.\ref{B0rPlot}. From the linear least-squares best fit, we found a single power-law falloff index, $\alpha_B=$  $-1.23 \pm 0.18$, which can represent the variation of B$_0$ from near-Sun to 0.99 AU.

\section{Discussion and conclusions}\label{discussions& Conclusions}


This work estimates the near-Sun axial magnetic field of a CME that occurred on 10 March 2022 from source NOAA AR 12962. We followed the methodology of \citet{Patsourakos_2016}, which relies on the magnetic helicity conservation principle. At the time of the eruption, the SolO spacecraft was located 7.8 degrees east of the Sun-Earth line, allowing for both in situ and remote sensing SolO observations at 0.43 AU. The WIND mission at L1 (0.99 AU) complements the in situ measurements, 
while STEREO and SOHO/LASCO contribute multi-viewpoint coronographic remote-sensing observations. We analysed the temporal evolution of the magnetic helicity of AR 12962 using the CB discrete flux tube method proposed by \citet{Georgoulis_2012}. Our estimations revealed a decrease in helicity within the AR region near CME onset, which is assumed to correspond to the helicity bodily transported to the CME. To estimate the near-Sun axial magnetic field of the CME, we employed the Lundquist flux rope model, utilising Eq.\ref{eqnof B estimtion} at a distance of 0.03 AU. We then extrapolated the near-Sun magnetic field using a power-law radial decrease  assumption 
at different distances (near-Sun, 0.43 AU, 0.99 AU) and estimated a single power-law index $\alpha_B$.

Our main findings are as follows:

\begin{enumerate}
\item The maximum helicity budget of the source AR in the pre-eruption phase, recorded on 10 March 2022, at 16:22 UT, is \((-9.94 \pm 1.2) \times 10^{41}\; \mathrm{Mx^2}\).
\item The net helicity difference in AR 12962 between the pre-and post-eruption phases is $ (-7.1 \pm 1.2) \times 10^{41}\; \mathrm{Mx^2} $. This represents approximately 71\%  of the total helicity budget, which is assumed to be fully transferred to the CME.
\item Utilising the Lundquist flux rope model, the axial magnetic field of the analysed CME at a distance of 0.03 AU is estimated at 2067 $\pm$ 405  nT.
\item Based on the power-law variation of $B_0$ with heliocentric distance, a single power-law index value, $\alpha_B$ from near-Sun-SolO-Earth is estimated to be $1.23 \pm 0.18$.
\item The estimated $\alpha_B$ from the near-Sun to SolO is $-1.17 \pm 0.14$, and from in situ measurements between SolO and WIND, it is $1.95 \pm 0.13$.
\end{enumerate}


We would like to compare the magnitude of our estimated CME helicity budget of $(-7.1 \pm 1.2) \times 10^{41}\; \mathrm{Mx^2}$ with previous studies on other CMEs/ARs.
Using the helicity injection method, \citet{Georgoulis2009ApJ}  studied solar magnetic helicity injected into the heliosphere over solar cycle 23 and found that the average helicity content per active-region CME varies between  $1.8 \times 10^{42}\;\mathrm{Mx^2}$ and $7\times 10^{42}\; \mathrm{Mx^2}$, with the upper extreme being an order of magnitude higher than ours. 
\citet{DeVore2000}, who studied magnetic helicity generation by solar differential rotation, employs the surface rate integral as a diagnostic tool to analyse the generation of helicity by horizontal motions of the footpoints of magnetic structures and finds that a typical interplanetary magnetic cloud contains a helicity of approximately $-2 \times 10^{42}\; \mathrm{Mx^2}$ and a magnetic flux of $\sim 10^{21} \mathrm{Mx}$, marking an order of magnitude higher than ours for helicity content. In \citet{Patsourakos_2016}, three methods were used to calculate the helicity of the CME, all indicating a net negative helicity, with smaller changes in positive helicity also observed. The helicity injection method \citep{paritat2006} yielded $-3.26 \times 10^{43} \mathrm{Mx^2}$, marking two orders of magnitude higher.
The CB method of \citet{Georgoulis_2012} had an estimate of $-4 \times 10^{42} \mathrm{Mx^2}$ and the volume calculation \citep{Moraitis2014} yielded $-8 \times 10^{42} \mathrm{Mx^2}$, both marking a order magnitude higher. However, the helicity injection method is expected to yield higher helicity values, as these values represent the total helicity injected into the system from the photospheric boundary and from the beginning of the observing sequence to the eruption onset. On the other hand, the CB method provides a lower-limit estimate of the CME's helicity because it assumes simple arch-like tubes and does not account for the braiding of flux tubes in the corona \citep{georgoulis2012}. A comparison of the CB method with several other helicity estimation methods, demonstrating its lower-limit estimations, has been discussed by \citet{Thalmann2021MagneticHE}. Furthermore, \citet{Tziotziou_2012}, using the connectivity matrix of the CB method, distinguished eruptive ARs from non-eruptive ones using both relative helicity and free magnetic energy, with helicity and free energy thresholds for the occurrence of major flares of $2 \times 10^{42} \, \mathrm{Mx^2}$ and $4 \times 10^{31} \, \mathrm{erg}$, respectively marking an order of magnitude higher than ours for helicity budget. Additionally, a recent study by \citet{Liokati_2022}, using helicity injection method, found thresholds for both the magnetic helicity and energy thresholds of $9 \times 10^{41} \, \mathrm{Mx^2}$ and $2 \times 10^{32} \, \mathrm{erg}$, respectively, which, if exceeded, suggest the host AR is likely to erupt. Estimates in this latter study are closer to our results.

In the era of PSP and SolO missions, co-aligned observations of eruptions such as the one studied here will provide stronger observational tests for 
estimating the axial magnetic field of Earth-directed CMEs. An example of such a coordinated observation occurred on 5 September 2022, when PSP observed from a heliocentric distance of 13.3 $R_{\odot}$ an event originating from AR 13088. 
This particular event has been extensively analysed by \citet{PSPSEVENTRomeo_2023}, \citet{PSPSEVENTDAVIS2023ApJ} and \citet{PSPSEVENTPaouris_2023}. PSP's recordings at 13.3 $R_{\odot}$ indicated a maximum ICME magnetic field strength of 1104 nT. A rough comparison between the estimated near-Sun magnetic field strength of 2067 $\pm$ 405 nT a distance of 6.6 $R_{\odot}$  of our event and the first-ever near-Sun in situ magnetic field measurement at 13.3 $R_{\odot}$ falls nearly at the same range.

In short, while variations exist, as the choice of methodology significantly influences the reported helicity values, our estimation of the helicity content agrees with previous studies within an order of magnitude. We tend to give a relatively low helicity content for the CME. This can also be attributed to our relatively weak AR source. It could also well be because of two reasons besides the weak source: first, the CB method provides a lower limit of helicity. Second, we assign the difference between the pre-and post-eruption phases to the CME rather than the entire helicity content of the source.


We now compare our estimated single power-law index from near-Sun to 0.99 AU ($\alpha_B = -1.23 \pm 0.18$) with previous studies that considered either near-Sun magnetic field estimations or measurements. \citet{Patsourakos_2016} found a value of $\sim -2$ for their study of the 2012 March 7 event using the same methodology, which is beyond the uncertainty limits of our result. However, this CME originated from a flux-laden AR with significantly higher helicity values and corresponded to an ultra-fast CME from NOAA AR 11429. On a follow-up statistical study of this methodology, in \citet{PatsourakosPrametric} used the input parameter distributions derived from observations to determine near-Sun and L1 magnetic fields for synthetic CMEs obtained $\alpha_B = -1.6 \pm 0.2$ , which marginally falls within our estimated range within the uncertainty range. This study has considered more flux rope models along with the Lundquist model. Furthermore, a recent study of \citet{salman2024PSPsurvey}, which presents a statistical investigation of the radial evolution of 28 ICMEs, measured in situ by PSP, at various heliocentric distances ranging from 0.23 to 0.83 AU, from October 2018 to August 2022. They considered ICME average magnetic field in contrast to the maximum value in our study and found an average value of  -1.21 $\pm$ 0.44 for $\alpha_B$, suggesting a less steep fall-off of the magnetic field with distance, which is in significant alignment with our results.

Additionally, we compare the obtained power-law index of -1.95 $\pm$ 0.13 from SolO-L1 obtained from in situ measurements in the inner heliosphere (r $<$ 1 AU) to other previous studies within the inner heliospheric region spanning 0.3 to 0.99 AU. Comparison with a study by \citet{Salman2020ExapnsionIndex}, which examined 47 ICMEs measured in situ by multiple spacecraft, studied the variation of magnetic sheath and ejecta separately; for the ejecta, the median was -1.75, with 50\% of the values falling within the range of -2.29 to -1.35. Additionally, in \citet{Self_similaritypowerlawGOOD2019}, a fitting analysis of 26 ensemble values of $B_0$ yielded a $\alpha_B = -1$.76 $\pm$ 0.04. Both of these studies fall within the uncertainty limits of our estimated value, indicating consistency in magnetic field variation trends. Conversely, a study by \citet{Leitner2007} for 130 events resulted in $\alpha_B = -1$.64 $\pm$ 0.40. This finding suggests a similar decreasing trend in $B_0$ with distance, though with a slightly different rate compared to \citet{Self_similaritypowerlawGOOD2019}. It also explores the individual radial decrease $B_0$ in each ICME by performing separate power-law fits to the two B$_0$ values obtained from each of the 13 events and obtained a mean fit $\alpha_B = -1$.34 $\pm$ 0.71. These $\alpha_B$ values have a slightly broader range of uncertainty, suggesting a potential variation in magnetic field behaviour within the inner heliosphere, though still overlapping with our results.  Similar work is found in  \citet{Farrugia2005ESASP}, where a study on the statistical variation of quantities associated with magnetic cloud propagation in inner heliospheric distances, utilising data from 31 events from Helios 1, 2 and Wind spacecraft data, concluded a $\alpha_B$ of -1.38 for the axial magnetic field. This estimation lies outside the uncertainty limits of our results. However, the absence of uncertainty reporting in \citet{Salman2020ExapnsionIndex} and \citet{Farrugia2005ESASP} studies poses challenges in definitive comparison.

In conclusion, the studies in the region 0.3-0.99 AU give rise to significant variability in $\alpha_B$ values. Our study contributes an important perspective with an estimated single $\alpha_B$ of -1.23 $\pm$ 0.18, encompassing observations from near- Sun to 0.43 AU up to  0.99 AU. The inclusion of the 0.43 AU data point offers a valuable validation opportunity for our methodology. Notably, our findings exhibit qualitative alignment with prior studies of \citet{PatsourakosPrametric}, which employed the same methodology for estimating near-Sun magnetic field characteristics for CME events. Successively, the alignment of our results with the statistical results of PSP measurements in \citet{salman2024PSPsurvey} is encouraging.
These findings also support the need for a more nuanced approach to modelling and predicting the dynamics of CMEs and ICMEs, considering their individual characteristics and the specific conditions of each event.


In summary, in addition to the estimation of the near-Sun magnetic field of this CME and an enhancement of understanding of the power-law variation of the CME magnetic field with heliocentric distance, our study confirms that the pre-and post-eruptive helicity difference in source ARs can be used for further study of the resulting CME. The current availability of multiple viewpoints and co-aligned observations of CME events allows the construction of a database intended to establish a maximum likelihood near-Sun CME $B_0$ that stems from observations rather than models. Our methodology has the potential to provide a foundation for routine calculations of magnetic helicity in the lower solar atmosphere combined with existing geometric modelling of CMEs in the outer corona. The creation of a near-Sun CME $B_0$ database and our methodology will hopefully contribute to a more systematic understanding of the near-Sun CME evolution. The maximum likelihood values of near-Sun CME $B_0$ and the understanding of its variation 
with heliocentric distance could also be used as initial conditions for MHD models simulating CME propagation in the inner heliosphere, such as the  European Heliospheric FORecasting Information Asset (EUHFORIA) modelling facility \citep{EUHFORIA2018}. 

\begin{acknowledgements}
This work is part of the SWATNet project funded by the European Union’s Horizon 2020 research and innovation programme under the Marie Skłodowska-Curie grant agreement No 955620. AN and SP acknowledge support from the ERC Synergy Grant ‘Whole Sun’ (GAN: 810218). The authors thank the referee for constructive comments/suggestions.

\end{acknowledgements}

\newpage
\bibliographystyle{aa-note} 
\bibliography{msref}


\begin{appendix}
\section{Data gap resolution of MAG/SolO in situ observations}\label{data gap MAG/Solo}
\begin{figure}[h!]
    \centering
    \includegraphics[width=1\linewidth]{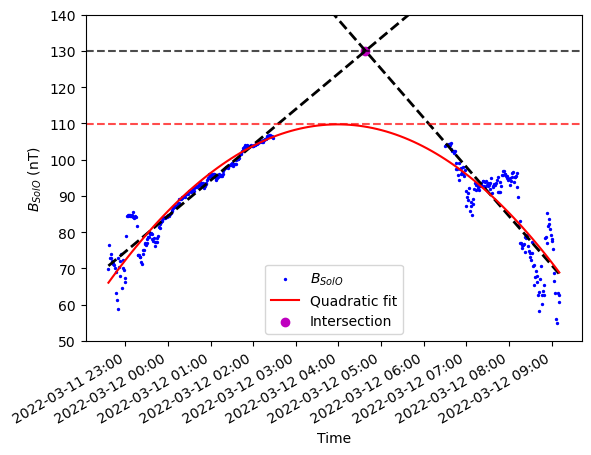}
    \caption{Scattered blue data points show in situ measurements of $B_{SolO}$ over time. The red curve represents the quadratic fit approximating the data trends. Extrapolated linear fits of $B_{SolO}$ nT and their intersection point (magenta dot) suggest potential $B_{0 SolO}$ at 0.43 AU.
}
    \label{datagapfitplot}
\end{figure}

We identified a significant data gap in SolO MAG and PAS measurements from 12-03-2022 02:30 UT to 12-03-2022 06:30UT due to instrumental problems. The low latency data published from these instruments didn't have the data gap which is used in some recent studies such as in \citet{Laker2023}. However, even after meticulous calibration procedures, it is noted that the L2 level normal 1-minute MAG data with the latest access on 14 March 2024 (the data used in this study) still exhibits a gap during the aforementioned period. This period encompasses the expected maximum magnetic field of our ICME.

To interpolate data in the gap, we used linear and quadratic fitting routines, as shown in Fig.\ref{datagapfitplot}: we selected the maximum magnetic field measurements ranging from 2022-03-11 22:36:00 UT to 2022-03-12 09:11:00. These time points correspond to the beginning and end of the period in which the magnetic field strength is anticipated to the peak. Firstly, we performed two separate linear fits to the data and then found the intersection between these lines. This value, found to be 130 nT, estimates the maximum anticipated magnetic field peak by at 0.43 AU. Secondly, a single quadratic fit was applied to all data points, yielding a peak of 109.72 nT. The discrepancy between the maximum anticipated peak (linear fits) and the quadratically inferred peak is considered as the uncertainty in the maximum magnetic field at SolO; hence, the  ICME maximum magnetic field measurement at 0.43 AU ($B_{0 SolO}$) by MAG/SolO is calculated as 109.72 $\pm$ 20.27 nT.

\section{Monte-carlo simulation for estimation of the near-Sun $B_0$ uncertainty}\label{monte-carlo procedure}

We employed a Monte Carlo simulation approach to calculate the axial magnetic field and its associated uncertainty. We compared the effects of uniform and normal distributions of input parameters on the results, focusing on the convergence and statistical reliability of the estimated axial magnetic field. However, our normal distribution simulations revealed a critical issue: a significant portion of the samples resulted in negative values for parameters such as $H_m$, $R$ or $L_{CME}$. These negative values are unphysical in the case of R or $L_{CME}$, leading to impossible estimates of $B_0$, that is, negative values inside the square root in Eq.\ref{eqnof B estimtion}. This caused us to discard samples with at least one negative input value, as indicated in Table \ref{montecarloNormal} and Table \ref{MontecarloUniform}. This process eliminated half of the samples in the normal distribution, undermining its statistical reliability and preventing the convergence of the estimated value of $B_0$. In contrast, assigning a uniform distribution that assigns equal probability to all values within an uncertainty range, which is useful when prior knowledge is limited, yielded no combinations with negative or unphysical inputs, as seen in Table \ref{MontecarloUniform}. This made the sampling statistically significant and showed improved convergence regardless of sample size.

Our analysis has revealed that selecting a probability distribution in Monte Carlo simulations can substantially affect the outcomes, particularly in systems sensitive to input variability. The normal distribution, which is often used, was found to be inadequate in our study due to its tendency to generate unrealistic scenarios. Therefore, it is important to be careful when selecting distributions, particularly when the physical meaning of inputs is essential. The uniform distribution within the Monte Carlo simulation framework was found to be more suitable for this context. It provided more precise estimations of $B_0$ and applicable uncertainty, avoiding convergence issues and producing realistic results. Additionally, the estimated uncertainty remained relatively consistent regardless of the sample size. We suggest that future research explore other distributions or hybrid approaches, particularly in situations with limited prior knowledge about the true distribution of input parameters.

\begin{table}
\caption{Results of Normal distribution trials}
\label{montecarloNormal}
\centering
\begin{tabular}{ccccc}
\hline
N.samples & N. Eliminated & B$_{0}$mean (Gauss) & B$_{0}$std (Gauss) \\
& combinations& & &\\
\hline
100 & 5 & 0.025 & 0.013 \\
500 & 23 & 0.023 & 0.017 \\
1000 & 175 & 0.024 & 0.022 \\
10000 & 4074 & 0.017 & 0.024 \\
\hline

\end{tabular}
\end{table}

\begin{table}
\caption{Results of uniform distribution trials}
\label{MontecarloUniform}
\centering
\begin{tabular}{ccccc}
\hline
N.samples & N. Eliminated & B$_{0}$mean (Gauss) & B$_{0}$std (Gauss) \\
& combinations& & &\\
\hline
100 & 0 & 0.021 & 0.004 \\
500 & 0 & 0.021 & 0.005 \\
1000 & 0 & 0.021 & 0.004 \\
10000 & 0 & 0.021 & 0.004 \\
\hline
\end{tabular}
\end{table}

\end{appendix}

\end{document}